\documentclass[12pt]{article}
\usepackage{graphicx}
\usepackage{caption}
\usepackage{subcaption}
\usepackage{amsmath,amssymb,mathrsfs,float,afterpage} 
\usepackage[longnamesfirst]{natbib}
\usepackage{tikz}

\begin{document}

\pagestyle{plain}

\newcommand{\spc}[1]{\Pisymbol{cryst}{#1}}
\newcommand{\spctf}[1]{\begin{turn}{45} \Pisymbol{cryst}{#1}
\end{turn}} 
\newcommand{\spctn}[1]{\begin{turn}{90} \Pisymbol{cryst}{#1}
\end{turn}}

\def\bA{\mbox{\boldmath$A$}}
\def\ba{\mbox{\boldmath$a$}}
\def\bB{\mbox{\boldmath$B$}}
\def\bb{\mbox{\boldmath$b$}}
\def\bC{\mbox{\boldmath$C$}}
\def\bc{\mbox{\boldmath$c$}}
\def\bD{\mbox{\boldmath$D$}}
\def\bd{\mbox{\boldmath$d$}}
\def\bE{\mbox{\boldmath$E$}}
\def\be{\mbox{\boldmath$e$}}
\def\bF{\mbox{\boldmath$F$}}
\def\boldf{\mbox{\boldmath$f$}}
\def\bG{\mbox{\boldmath$G$}}
\def\bH{\mbox{\boldmath$H$}}
\def\bh{\mbox{\boldmath$h$}}
\def\bI{\mbox{\boldmath$I$}}
\def\bJ{\mbox{\boldmath$J$}}
\def\bK{\mbox{\boldmath$K$}}
\def\bk{\mbox{\boldmath$k$}}
\def\bL{\mbox{\boldmath$L$}}
\def\bM{\mbox{\boldmath$M$}}
\def\bN{\mbox{\boldmath$N$}}
\def\bn{\mbox{\boldmath$n$}}
\def\bP{\mbox{\boldmath$P$}}
\def\bp{\mbox{\boldmath$p$}}
\def\bq{\mbox{\boldmath$q$}}
\def\bR{\mbox{\boldmath$R$}}
\def\br{\mbox{\boldmath$r$}}
\def\bS{\mbox{\boldmath$S$}}
\def\bT{\mbox{\boldmath$T$}}
\def\bt{\mbox{\boldmath$t$}}
\def\bU{\mbox{\boldmath$U$}}
\def\bu{\mbox{\boldmath$u$}}
\def\bv{\mbox{\boldmath$v$}}
\def\bW{\mbox{\boldmath$W$}}
\def\bw{\mbox{\boldmath$w$}}
\def\bX{\mbox{\boldmath$X$}}
\def\bx{\mbox{\boldmath$x$}}
\def\by{\mbox{\boldmath$y$}}
\def\bZ{\mbox{\boldmath$Z$}}
\def\bz{\mbox{\boldmath$z$}}

\def\bbeta{\mbox{\boldmath$\beta$}}
\def\beps{\mbox{\boldmath$\epsilon$}}
\def\bSigma{\mbox{\boldmath $\Sigma$}}
\def\btau{\mbox{\boldmath $\tau$}}
\def\btheta{\mbox{\boldmath $\theta$}}
\def\blam{\mbox{\boldmath $\lambda$}}

\def\cc{{\cal C}}
\def\calr{{\cal R}}
\def\cw{{\cal W}}
\def\cx{{\cal X}}
\def\cz{{\cal Z}}

\def\bzero{\mbox{\boldmath $0$}}
\def\b1{\mbox{\boldmath $1$}}
\def\bsim{\mbox{\boldmath $\sim$}}

\def \ni{\noindent}
\def \ds{\displaystyle}
\def \ul{\underline}
\def \fns{\footnotesize}
\def \ds{\displaystyle}

\newcommand{\hg}[2]{\mbox{}_{\scriptscriptstyle #1} F_{\scriptscriptstyle #2}}
\def \hlam{\hat{\lam}}
\def \hp{\hat{p}}
\newcommand{\overbar}[1]{\mkern 1.5mu\overline{\mkern-1.5mu#1\mkern-1.5mu}\mkern 1.5mu}

\def \mur{\frac{\mu}{\mu+r}}
\def \rmu{\frac{r}{\mu+r}}

\def \a{\alpha}
\def \b{\beta}
\def \s{\sigma}
\def \e{\epsilon}
\def \ul{\underline}
\def \t{\theta}
\def \ds{\displaystyle}
\def \d{\delta}
\def \g{\gamma}
\def \lam{\lambda}
\def \om{\omega}
\begin{center}
{\bf Stereographic Projections for Designs on the Sphere}\\
Linda M. Haines\\
{\small Department of Statistical Sciences,
University of Cape Town, Rondebosch 7700, South Africa.\\
July 4, 2024}
\end{center}

\noindent{\bf Abstract}

This paper is concerned with the use of the stereographic projection to map the  points of a design on the sphere in three dimensions  onto a two-dimensional stereogram. Details of the projection and its attendant stereogram are given and the application of the constructs to designs on the sphere is highlighted. Spherical $t$-designs are then introduced and stereograms of selected examples are presented. In addition, examples of regression models with design space the unit ball are considered, stereograms which represent the spherical component of such designs are given and an example in which stereograms are used to elucidate the geometric isomorphism of a suite of designs generated by an exchange algorithm is presented. Emphasis throughout the text is placed on the use of the stereographic projection in  teaching and research within the framework of statistics.

\bigskip

\noindent{\bf 1. Introduction}

The aim of the present study  is to introduce the stereographic projection as a valuable tool in teaching and research for representing the points  of a design over the three-dimensional sphere as a stereogram in two dimensions. The projection is used extensively in, for example, crystallography, complex analysis and  geology but  has received scant attention in the statistics and related literature. The designs of interest here arise in areas as diverse as the distribution of charged particles on the sphere,  the nature and application of spherical $t$-designs and  regression models with design space the unit sphere.  \bigskip

The paper is organized as follows. The notion of the stereographic projection and the construction of the attendant stereogram are presented  in Section 2. Spherical $t$-designs are introduced in Section 3 and stereograms of selected examples are presented there. In Section 4 an example of a quadratic regression model in three variables, taken from the statistical literature, is discussed and the use of stereograms  to elucidate the geometric isomorphism of a suite of designs generated from the model by an exchange algorithm is presented. Some general comments  are included  in Section 5. The Sch\"{o}nflies notation for points groups is used to describe the structures of the designs of interest formally. An R package for constructing the stereograms of the designs is also available from the author on request. 
\bigskip
\bigskip

\noindent{\bf 2. The Stereographic Projection}

\begin{figure}[htb!]
\vspace{-15.5mm}
\centering
\begin{subfigure}[b]{0.495\textwidth}
\includegraphics[width=\textwidth]{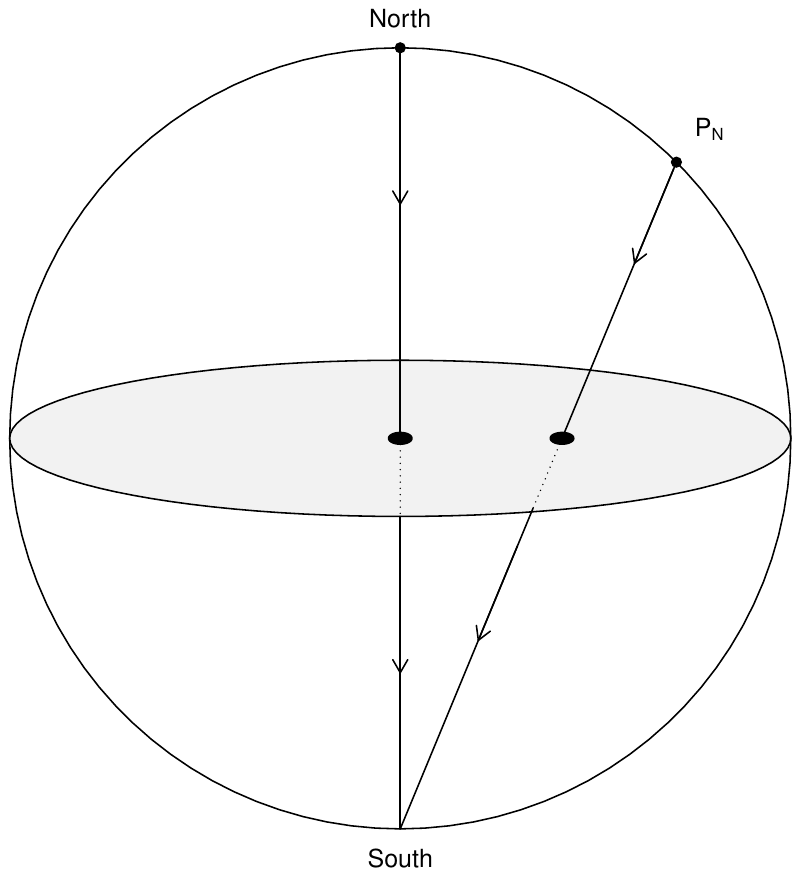}
\vspace{-35mm} \caption{Northern Hemisphere.}
\end{subfigure}
\hfill
\begin{subfigure}[b]{0.495\textwidth}
\includegraphics[width=\textwidth]{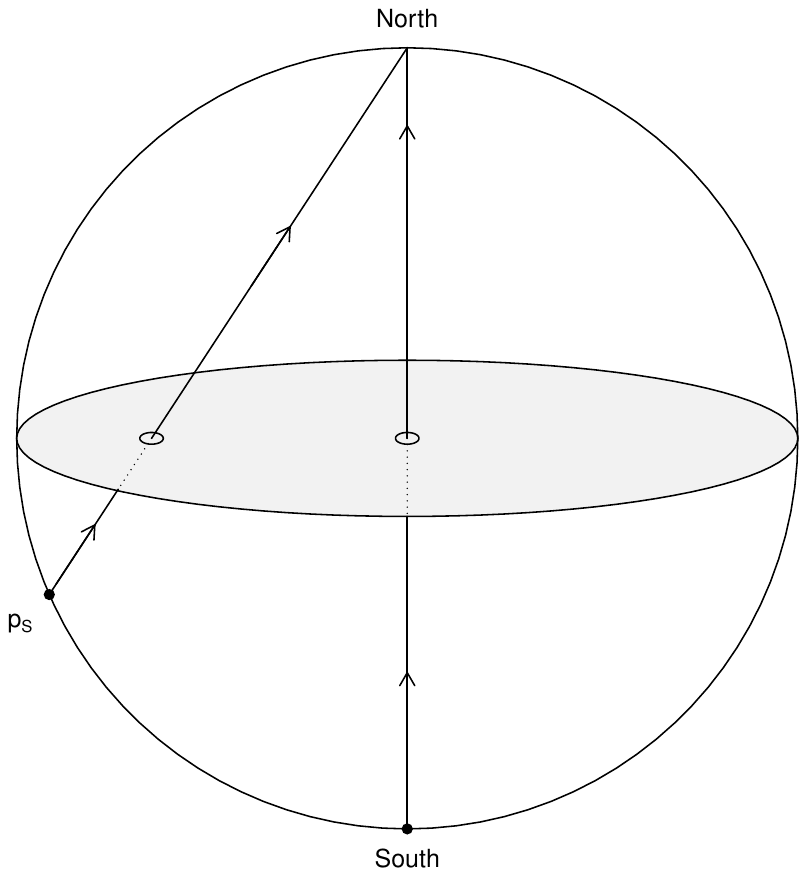} 
\vspace{-35mm} \caption{Southern Hemisphere.}
\end{subfigure}
\begin{subfigure}[b]{0.495\textwidth}
 \includegraphics[width=\textwidth]{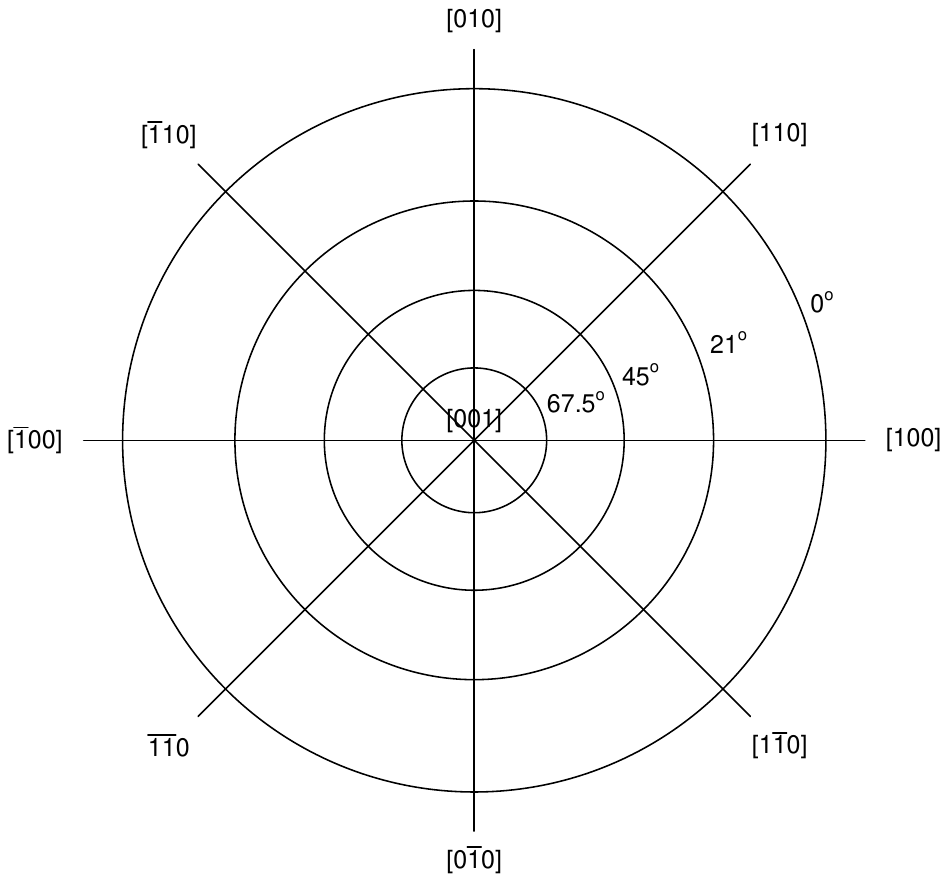} 
\vspace{-35mm}\caption{Meridians and Parallels.}
\end{subfigure}
\hfill
\begin{subfigure}[b]{0.495\textwidth}
 \includegraphics[width=\textwidth]{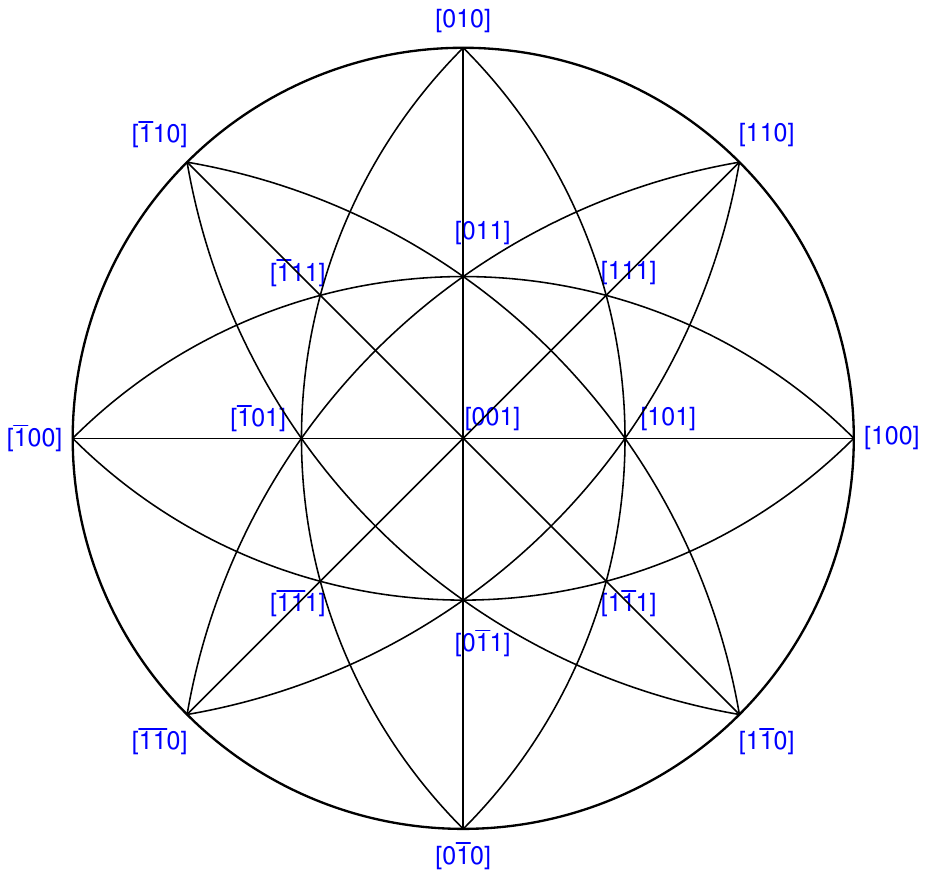} 
\vspace{-35mm} \caption{Meridians and Great Circles.}
\end{subfigure}
\vspace{-5mm}

\caption{Stereographic projections for points (a) in the northern hemisphere and (b) in the southern hemisphere. Stereograms, looking down the North-South axis,  of (c) meridians and parallels  and (d) meridians and great circles.}
\label{fig1}
\end{figure}

Consider first a perfect terrestrial globe which is centred at the origin with the North-South axis oriented vertically and the equatorial plane  perpendicular to that axis. The stereographic projection maps points and circles on the three-dimensional globe onto the two-dimensional equatorial plane, with the resulting construct termed a stereogram. In particular, consider a point lying on the surface of the globe in the northern hemisphere. Then a line can be drawn from that point to the South Pole and the   point of intersection of the line with the equatorial plane indicated on the stereogram with a solid circle, as shown in Figure \ref{fig1}(a). Similarly, a line can be drawn from a point on the surface of the globe in the southern hemisphere to the North Pole and its intersection with the equatorial plane denoted, for clarity, by an open circle, as in Figure \ref{fig1}(b).  
In addition, parallels, that is lines of latitude, are projected onto the equatorial plane as circles centred at the origin and meridians, that is lines of longitude, as straight lines through the origin. These constructs can be represented by a polar net, as shown in Figure \ref{fig1}(c). Great circles through two points on the surface of the sphere can also be  projected onto the equatorial plane as arcs  with diagonally opposite end points touching the equator. Selected meridians and great circles  are shown on the stereogram in Figure \ref{fig1}(d). 
Three-dimensional directions are often indicated on the stereogram and, for example, the vectors $(1,1,1)$ and $(-1,1,0)$ are denoted compactly by $[111]$ and $[\bar{1}10]$. To give a sense of the orientation of the equatorial plane, selected directions are included in Figure \ref{fig1}(d). Finally, note that a small circle drawn on the surface of the globe or sphere can be projected onto the stereogram  as a circle  but the centres of the two circles do not necessarily correspond. Note also that the stereographic projection preserves angles but not distances on the sphere.  A point group in three dimensions, that is a set of symmetry operations which leave at least one point unchanged, forms an integral part of the stereographic projection, with  the unchanged point taken to lie at the centre of the sphere.  In addition, orbits are very often used to highlight the symmetries which define a point group. Specifically, an orbit is a set of points which can be obtained by a group action of the point group on a given point in that set.
Point groups are presented in the Sch\"onflies notation throughout the text. 
Further details on the stereographic projection are given, for example, in the books on crystallography by  \citet{borch:11}, \citet{degm:12} and \citet{hammond:15}. Finally, it is worth noting that the exponential map, which can also be invoked to represent points on a sphere in two dimensions, preserves distances but not angles and is not relevant in the present context.

\bigskip

The stereographic projection can immediately be adapted to accommodate designs in three dimensions  over  the unit sphere.
Specifically, consider the perfect terrestrial globe described above as a unit sphere with a right-handed coordinate system specified by the coordinates $x_1, x_2$ and $x_3$. Then the point $(0,0,1)$ can be taken taken to coincide with the North Pole, the point $(0,1,0)$  with the intersection of a line in the direction $[010]$ with the equator and the $(x_1,x_2)$-plane with the equatorial plane. The support points of a design on the surface of the sphere can be projected onto the equatorial plane, that is the stereogram, by invoking the mapping $(x_1,x_2,x_3) \mapsto  \big( \displaystyle \frac{x_1}{1+|x_3|},\frac{x_2}{1+|x_3|} \big)$. Projections onto the stereogram  of  the appropriate meridians and great circles which connect the design points then follow. 

\bigskip
\bigskip

\noindent{\bf 3. Spherical $t$-Designs}

Spherical $t$-designs were  introduced into the literature on algebraic combinatorics by \citet{dgs:77} and are of considerable current interest both theoretically and practically. 
In the present context of three dimensions, the designs comprise points on the sphere  for which the average evaluated at a homogeneous spherical polynomial of degree $t$ is exactly equal to the integral of the polynomial with respect to the uniform distribution on the sphere. More formally, a spherical $t$-design is a finite subset $Y$ on the unit sphere $S^{2} \subset \mathbb{R}^{3}$ which satisfies the cubature relationship 
\begin{equation}
\int_{S^{2}}   f(x) d\sigma(x) = \frac{1}{|Y|} \sum_{y \in Y} f(y) ~~ \mbox{for~all~} f(x) \in \mbox{Pol}_t(S^{2}),
\label{spht}
\end{equation}
where $\sigma$ is a uniform measure on the sphere and $\mbox{Pol}_t(S^{2})$ denotes the space of homogeneous spherical polynomials of degree less than or equal to $t$.  This definition can be modified, following \citet{gp:11} and \citet[p.~14]{sawahk:19}, by replacing the  sum on the right-hand of Equation (\ref{spht}) with a weighted sum 
$\sum_{y \in Y} w_y f(y)$, 
where $w_y$ represents a weight $0 \le w_y < 1$ associated with $y \in Y$ such that $\sum_{y \in Y} w_y =1$. The resultant designs are referred to in the present context as weighted spherical $t$-designs. 
\bigskip

A number of examples of spherical $t$-designs and their weighted counterparts are now introduced in order to highlight the use of the stereogram in identifying  the nature and structure of the designs. Spherical $t$-designs in three dimensions are, in general,  not straightforward to construct  and the computational problems are highlighted in the papers by \citet{hs:96}  and  \citet{gp:11}. The examples presented here are drawn in digital form from Neil Sloane's website \citet{nsweb} for the spherical $t$-designs and from the website of \citet{graefweb} for the weighted spherical $t$-designs.
\bigskip

\begin{figure}
\centering
\vspace{-22.50mm}
\begin{subfigure}[b]{0.475\textwidth}
\includegraphics[width=\textwidth]{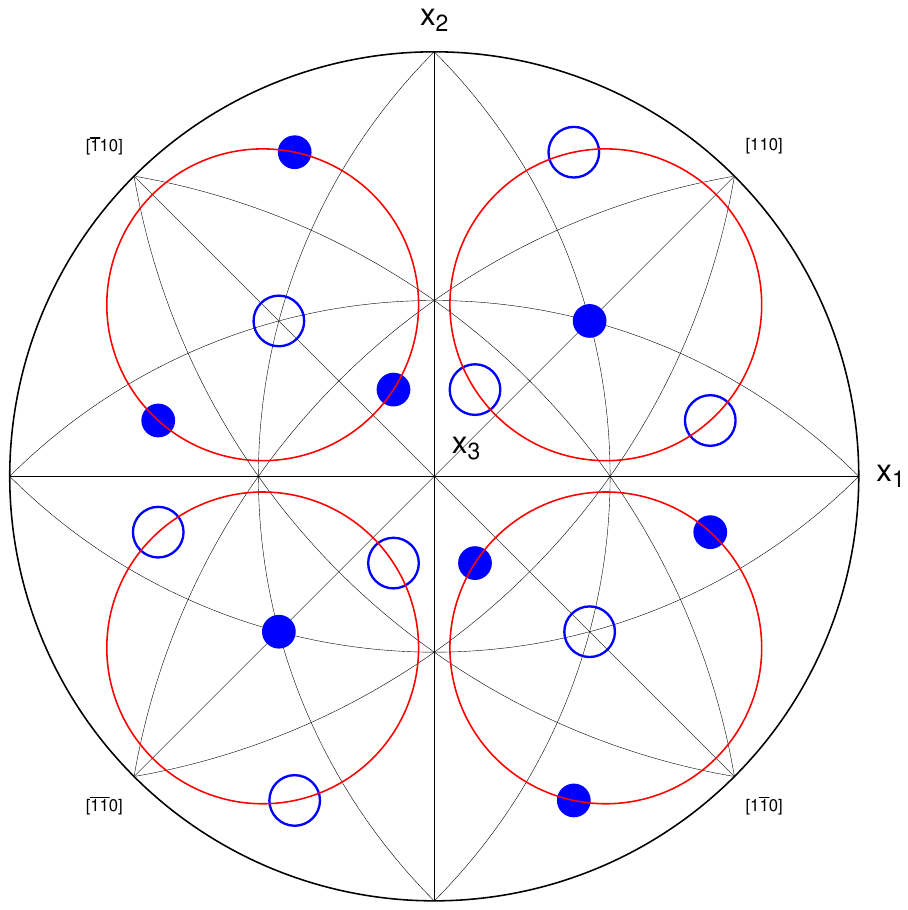}
\vspace{-35mm} \caption{16-point spherical 5-design, T.}
\end{subfigure}
\hfill
\begin{subfigure}[b]{0.475\textwidth}
\includegraphics[width=\textwidth]{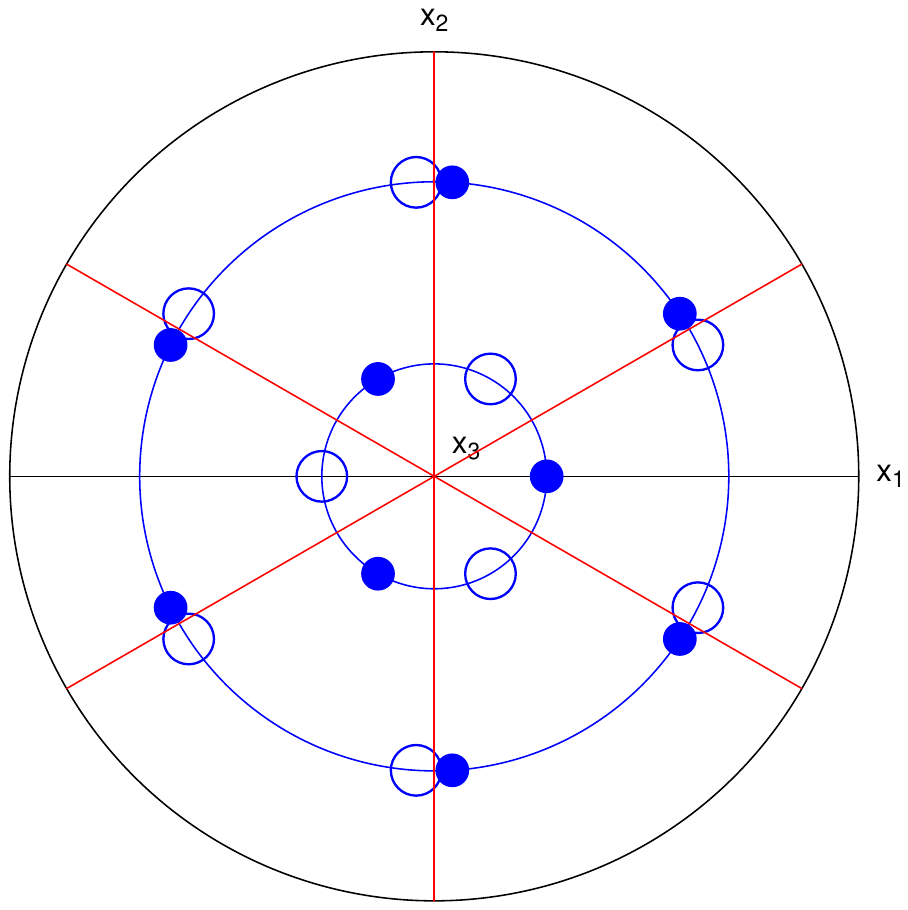} 
\vspace{-35mm} \caption{18-point spherical 5-design, D$_{3d}$.}
\end{subfigure}
\hfill
\begin{subfigure}[b]{0.475\textwidth}
\includegraphics[width=\textwidth]{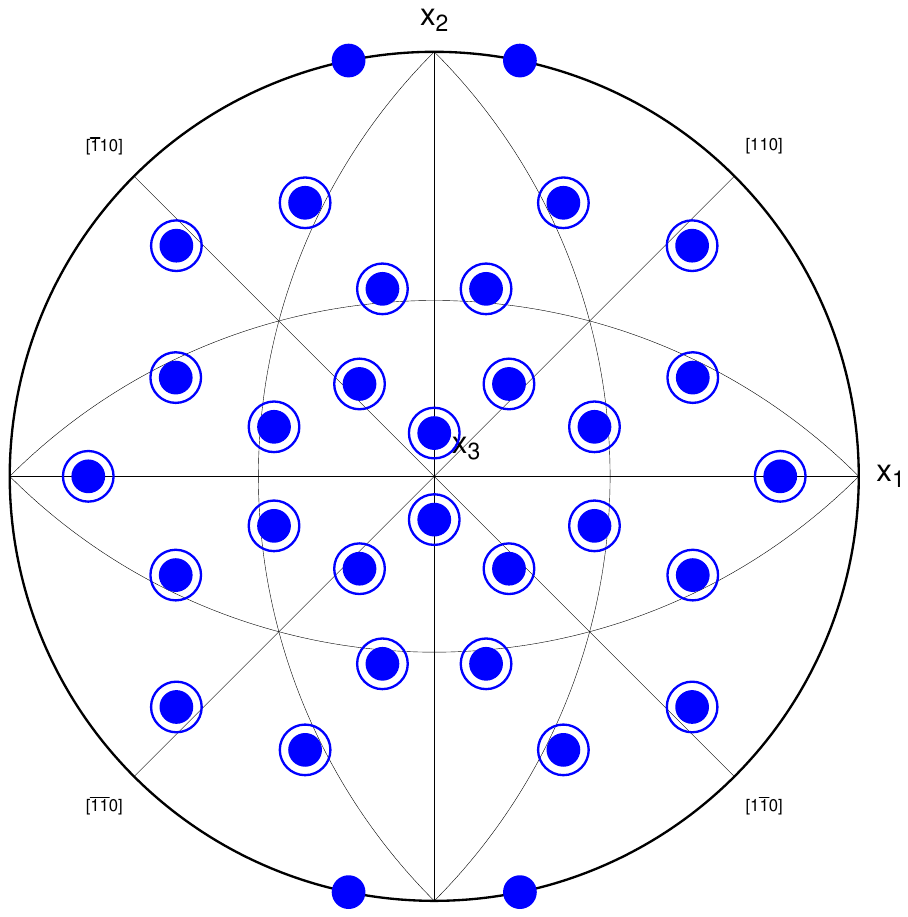} 
\vspace{-35mm} \caption{60-point spherical 5-design, I$_h$.}
\end{subfigure}
\hfill
\begin{subfigure}[b]{0.475\textwidth}
\includegraphics[width=\textwidth]{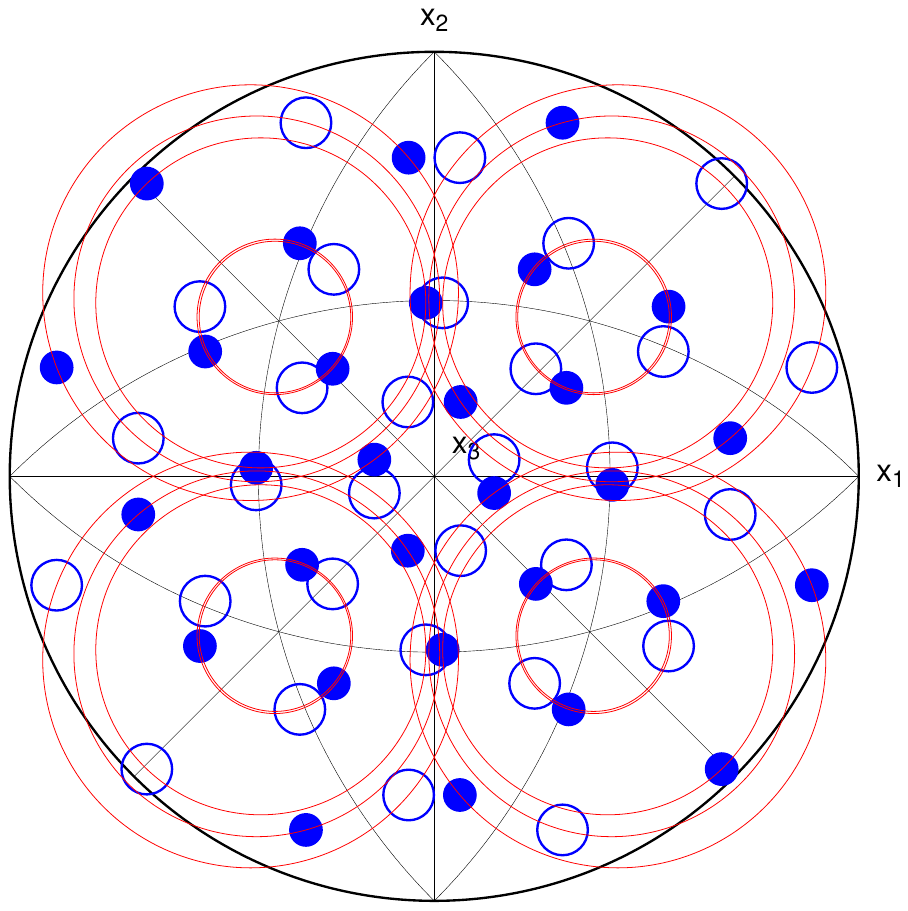} 
\vspace{-35mm} \caption{60-point spherical 10-design, T.}
\end{subfigure}
\vspace{10mm}

\caption{Stereograms (a) and (b) represent the 16-point and 18-point spherical 5-designs, with the snub tetrahedra in (a) and the trigonal axes in (b) highlighted in red  as circles and lines, respectively. The stereograms (c) and (d) represent the 60-point spherical $t$-designs with $t=5$ and $t=10$, respectively, with the five snub tetrahedra in (d) shown as red circles, the inner two of which are very close together.}
\label{fig2}
\end{figure}

\noindent{\bf Example 3.1: Spherical $t$-designs:}
Stereograms of $16$- and $18$-point spherical 5-designs and  60-point spherical $t$-designs with $t=5$ and $10$ are given in Figures \ref{fig2}(a), (b), (c) and (d), respectively, with  the appropriate point groups represented in Sch\"onflies notation in the captions. The $16$-point and $18$-point designs have the same $t$ values but very different structures. Specifically, it is clear from the stereograms that the $16$-point designs has tetrahedral symmetry, T, while the $18$-point design has a five-fold principal axis and symmetry D$_{5d}$. The two $60$-point designs have the same number of points but $t$ values of $5$ and $10$ and are structurally very different. Thus, as shown in stereogram \ref{fig2}(c), the $60$-point design with $t=5$ has icosahedral symmetry with point group I$_h$ and represents a football with hexagonal and pentagonal faces connected by edges. As an aside, it also represents the carbon allotrope, the Buckminsterfullerene, and its semi-spheres are geodesic domes. In contrast, the $60$-point spherical $10$-design shown in stereogram \ref{fig2}(d) comprises five snub tetrahedra and exhibits tetrahedral symmetry with point group T. 
\bigskip

\noindent{\bf Example 3.2: Weighted spherical $t$-designs}

\begin{figure}
\centering
\vspace{-40mm}
\begin{subfigure}[b]{0.475\textwidth}
\includegraphics[width=\textwidth]{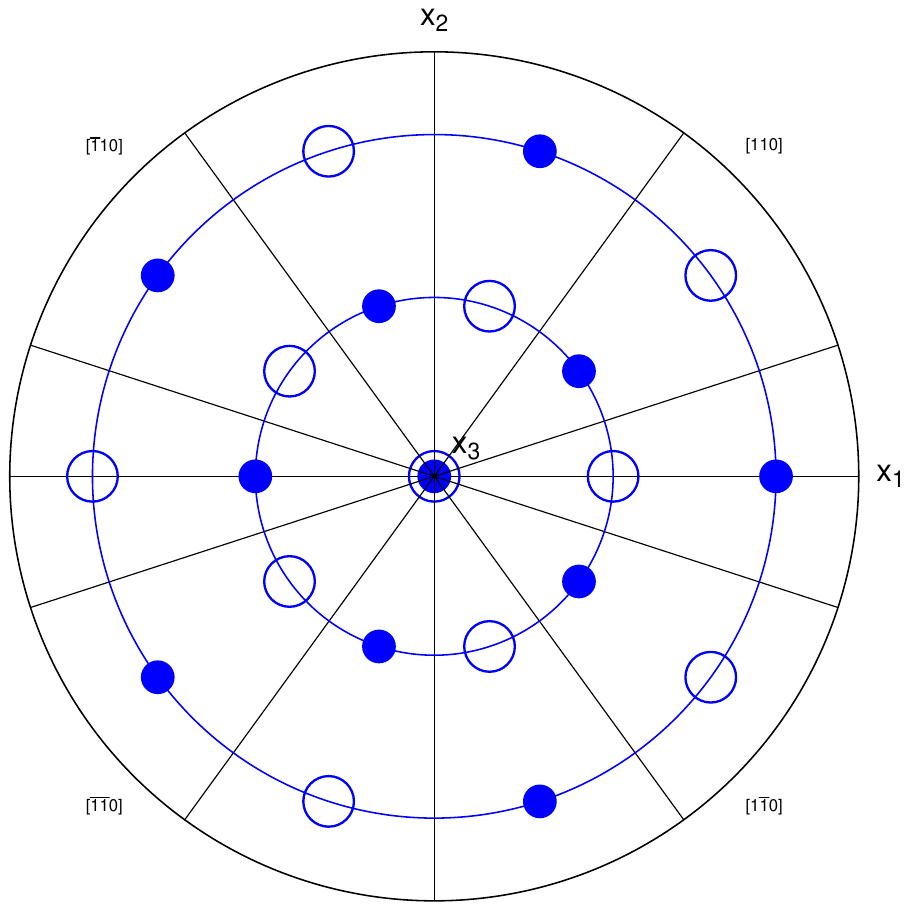}
\vspace{-35mm} \caption{22-point spherical 5-design, D$_{5d}$.}
\end{subfigure}
\hfill
\begin{subfigure}[b]{0.475\textwidth}
\includegraphics[width=\textwidth]{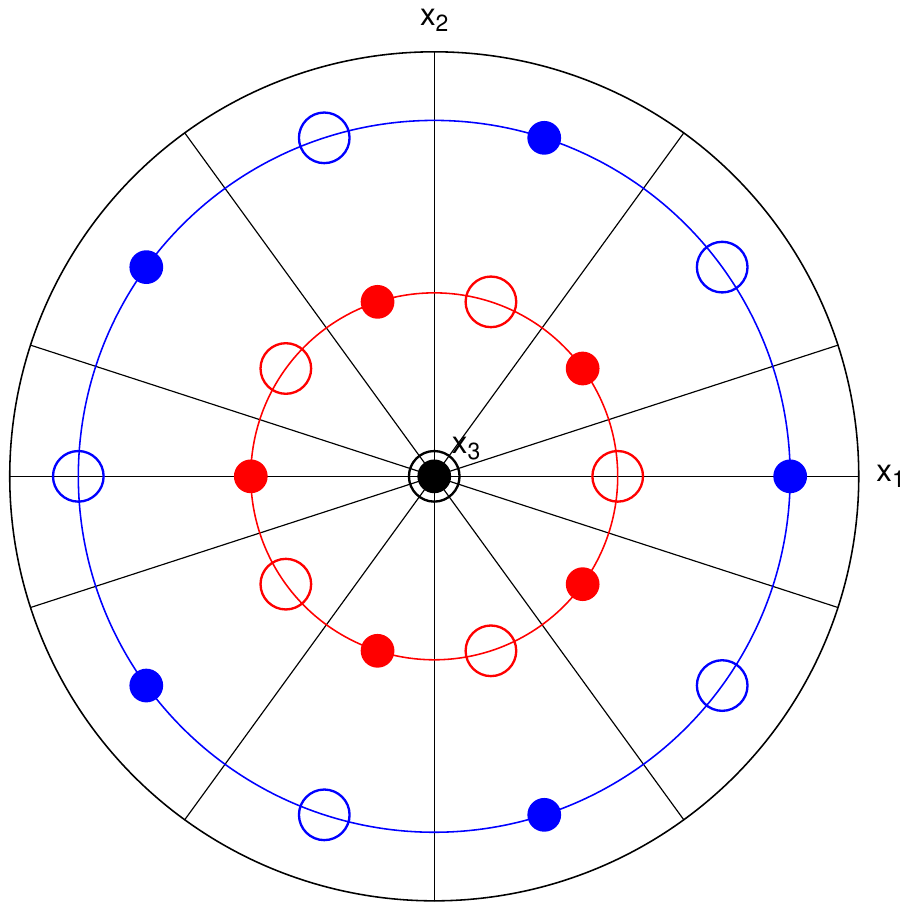} 
\vspace{-35mm} \caption{weighted 22-point  spherical 7-design, D$_{5d}$.}

\end{subfigure}
\hfill
\begin{subfigure}[b]{0.475\textwidth}
\includegraphics[width=\textwidth]{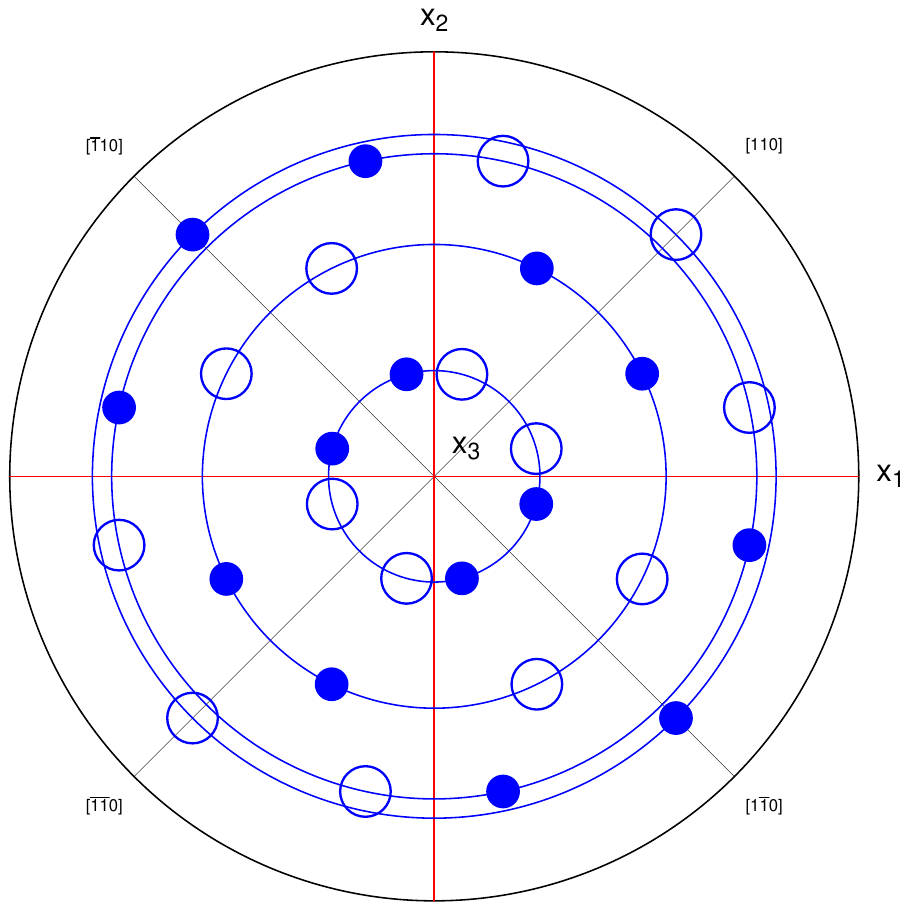} 
\vspace{-35mm} \caption{28-point spherical 6-design, D$_{2d}$.}
\end{subfigure}
\hfill
\begin{subfigure}[b]{0.475\textwidth}
\includegraphics[width=\textwidth]{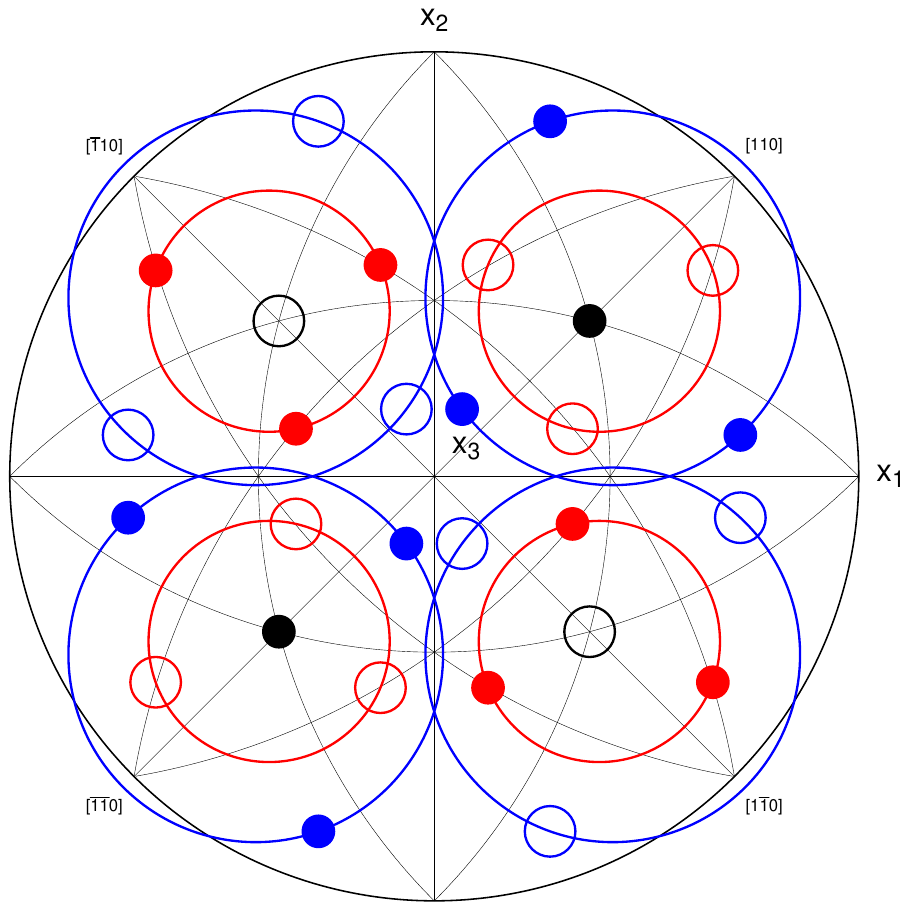} 
\vspace{-35mm} \caption{weighted 28-point  spherical 8-design, T.}
\end{subfigure}

\vspace{5mm}

\caption{Stereograms (a) and (b) represent the unweighted and weighted 22-point spherical $t$-designs with the orbits of the weighted  design indicated in black, red and blue with weights $0.0926, 0.4221$ and $0.4853$, respectively. Stereograms (c) and (d) represent the unweighted and weighted 28-point spherical $t$ designs, with the two-fold axes of the unweighted design highlighted in red and the orbits at and around the vertices of the tetrahedron of the weighted design in black, red and blue with weights $0.1385, 0.4537$ and $0.4078$, respectively.}
\label{fig3}
\end{figure}

Stereograms of the 22-point spherical 5-design and the weighted 22-point spherical $7$-design are shown in Figures \ref{fig3}(a) and \ref{fig3}(b).  The two designs clearly belong to the same point group, that is D$_{5d}$, and comprise orbits defined by the North and South Poles and by points located on two sets of north and south latitudes which are at the same polar angle from the equator. The designs do however exhibit subtle differences.  First, it is easily shown that the orbits comprising points located on the latitudes of the two designs are defined by different polar angles and that the latitudes associated with the weighted spherical $7$-design are closer to the equator. Second, and more importantly, the spherical 5-design comprises equally weighted points while, by definition, the weighted spherical $7$-design does not. Specifically, the orbits of the latter design are indicated in black, red and blue in Figure \ref{fig3}(b) and have  weights $0.0926, 0.4221$ and $0.4853$, respectively, that is  taken from the centre outwards.  
Stereograms of the 28-point spherical 6-design and its counterpart, the weighted 28-point spherical 8-design, are given in Figures \ref{fig3}(c) and \ref{fig3}(d). In this case, the two designs exhibit different symmetries, with the spherical $6$-design belonging to the point group D$_{2d}$ and the spherical $8$-design to the point group T. The orbits of the latter design are defined by the four points corresponding to the vertices of the tetrahedron and two sets of points located around the three-fold axes. The orbits are highlighted on the stereogram in Figure \ref{fig3}(d) in black, red and blue and have weights $0.1385, 0.4537$ and $0.4078$, that is from the vertices of the tetrahedron outwards, respectively.
\bigskip
\bigskip

\noindent{\bf 4. Regression Modelling}

Designs over  the unit ball which comprise a finite set of support points located at the centre of the ball and on the sphere have been reported sporadically in the literature as, for example, in the  papers of \citet{bd:59}, \citet{hs:91}, \citet{sg:06}, \citet{detteg:14}, \citet{hsj:15} and  \citet{rs:23}. Three-dimensional prototypes of these designs help to convey the innate structure of the  hyper-spherical designs and can be represented to some advantage  by invoking the stereographic projection and attendant stereogram.
\bigskip

\noindent{\bf 4.1 Pedagogical Examples}

To fix ideas, consider the subset designs for the quadratic polynomial model with three  independent variables over the unit ball introduced into the literature in  the paper by \citet{sg:06}. The support points of the designs lie on the surface and at the centre of the unit ball and are taken from the  sets specified by 
$$
S_0=(0,0,0), ~S_1=(\pm1,0,0)_{\mathrm{ \small cycle}}, ~S_2=(\pm \frac{1}{\sqrt{2}},\pm \frac{1}{\sqrt{2}},0)_{\mathrm{ \small cycle}},  ~S_3=(\pm \frac{1}{\sqrt{3}},\pm \frac{1}{\sqrt{3}},\pm \frac{1}{\sqrt{3}})
$$
which comprise $1, 6, 12$ and $8$ points, respectively.  The subscript `cycle' indicates that the three coordinates of the points can be moved cyclically.
More specifically, subset designs can  be expressed as a linear combination of these sets of points, that is as $c_0 S_0 +c_1 S_1+c_2 S_2 + c_3 S_3$, where $c_0 >0$ and the remaining coefficients, $c_1, c_2$ and $c_3$, are  greater than or equal to zero and are chosen in such a way that the moment matrix is nonsingular. From a symmetry perspective, the sets of points $S_0, S_1, S_2$ and $S_3$ correspond to the centre point and, for a cube inscribed in the unit sphere,  to the intersection of the six normals to the faces of the cube with the sphere, to the intersection of the 12 lines from the origin through the mid-points of the edges of the cube with the sphere, and to the eight vertices of the cube itself, respectively. It thus follows that the points on the unit sphere of a subset design with equal weights assigned to the subsets $S_1, S_2$ and $S_3$ belong to the point group $O_h$.
\bigskip

\begin{figure}[htb!]
\vspace{-22.50mm}
\centering

\begin{subfigure}[b]{0.495\textwidth}
\includegraphics[width=\textwidth]{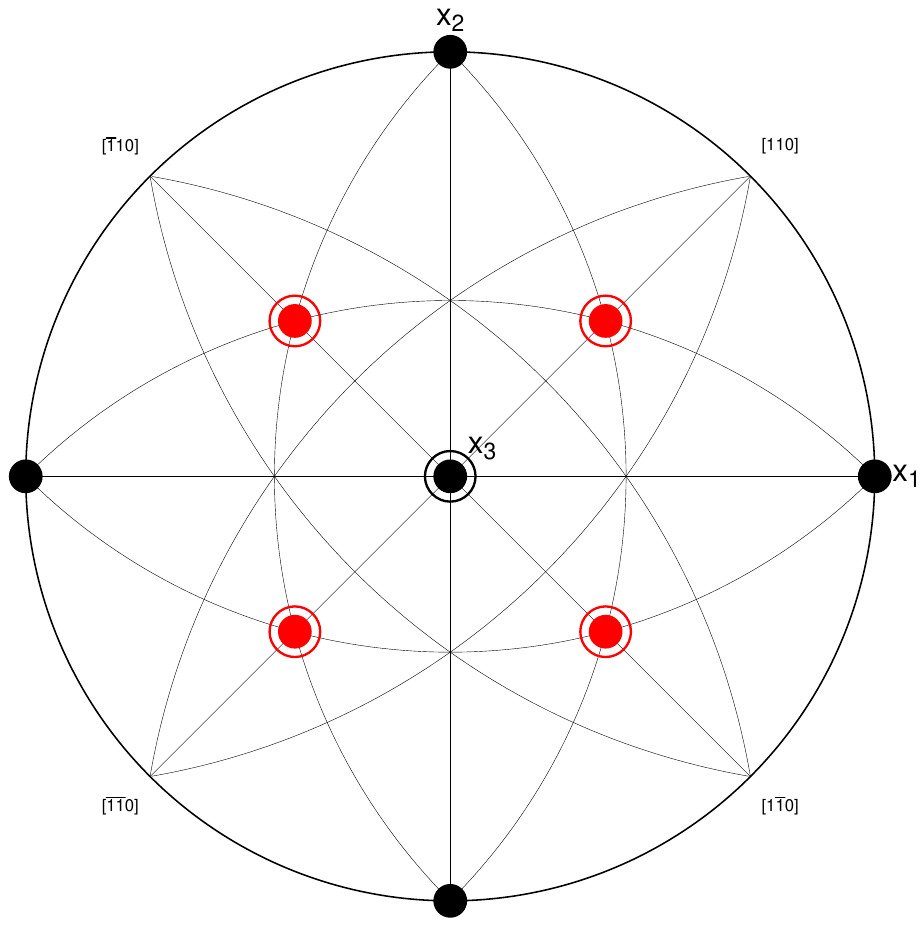}
\vspace{-35mm} \caption{Central Composite Design}
\end{subfigure}
\hfill
\begin{subfigure}[b]{0.495\textwidth}
\includegraphics[width=\textwidth]{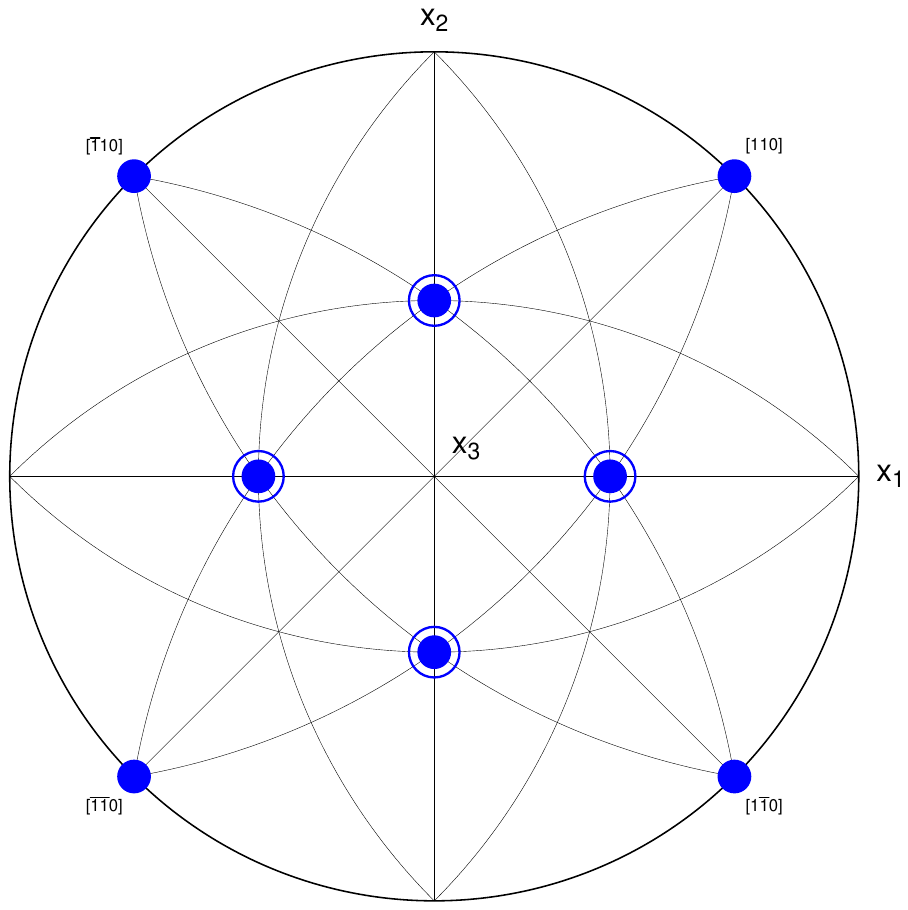} 
\vspace{-35mm} \caption{Box-Benkhen Design}
\end{subfigure}
\begin{subfigure}[b]{0.495\textwidth}
 \includegraphics[width=\textwidth]{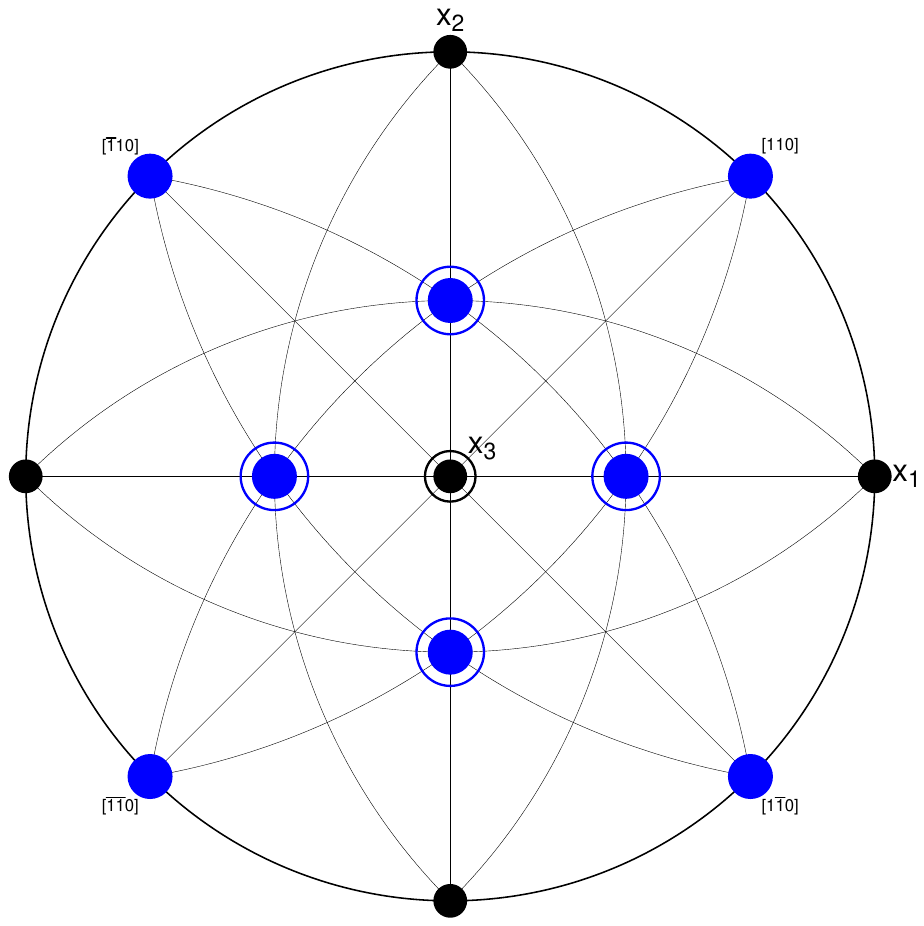} 
\vspace{-35mm}\caption{Rotatable Design}
\end{subfigure}
\hfill
\begin{subfigure}[b]{0.495\textwidth}
 \includegraphics[width=\textwidth]{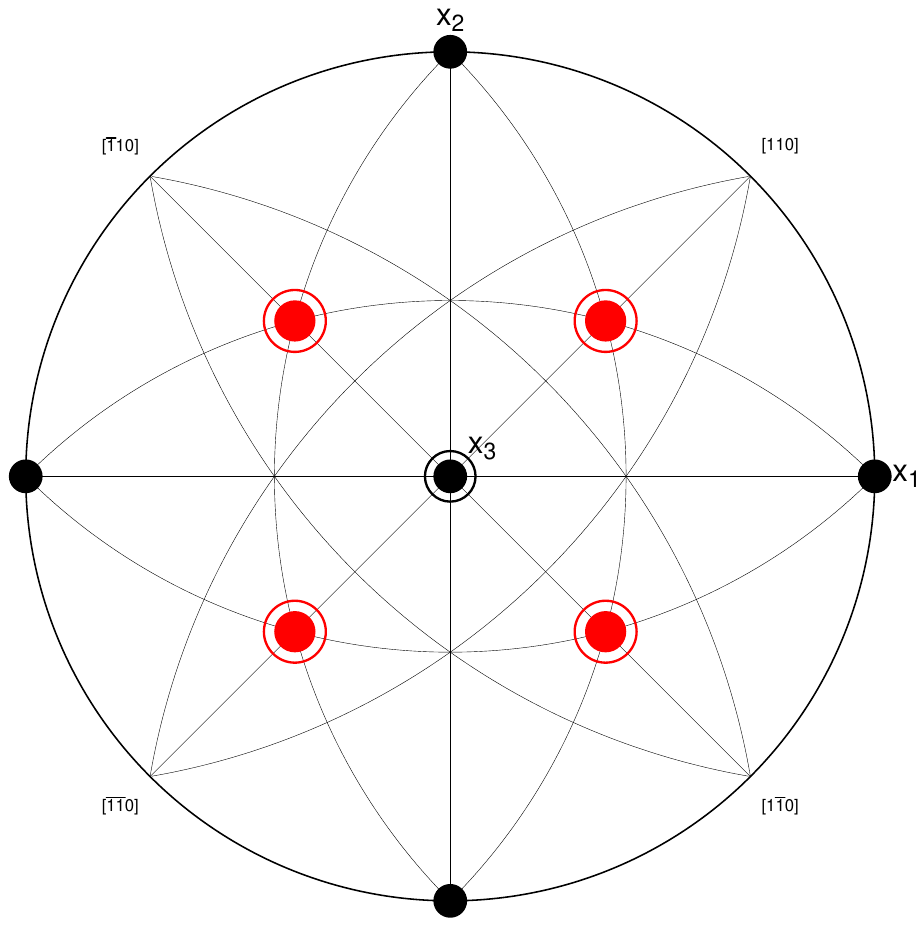} 
\vspace{-35mm} \caption{Approximate E-optimal Design}
\end{subfigure}
\vspace{-5mm}

\caption{Stereograms of the sphere points of (a) the central composite design, (b) the Box-Behnken design, (c) the rotatable design $S_0 + S_1 + 2 S_2$ and (d) the approximate $E$-optimal design $\frac{9}{17} S_0 + \frac{3}{17} S_1 + \frac{5}{17} S_3$. The points of the sets $S_1, S_2$ and $S_3$ are highlighted in black, blue and red. In the stereograms (c) and (d), the areas of the circles representing the points are proportional to the relative weightings of those points on the sphere.}
\label{fig4}
\end{figure}

The central composite design over design space the unit ball is the subset design $S_0+S_1+S_3$  and the stereographic projection of the 14 design points on the sphere yields the stereogram shown in Figure \ref{fig4}(a).
Similarly, the Box-Behnken design is the subset design $S_0 +S_2$ and the stereogram of the 12 points on the sphere is shown in Figure \ref{fig4}(b). Note that the points in $S_1, S_2$ and $S_3$ are coloured in the stereograms in black, blue and red in Figure \ref{fig4}, respectively, and that the complementary relationship between the two designs is clearly illustrated by the stereograms. Subset designs with different coefficients allocated to the sets of points $S_0, S_1, S_2$ and $S_3$ can also be represented compactly on a stereogram by using colour coding and points with areas proportional to their relative weightings. For example, the subset design $S_0+S_1+2 S_2$ is rotatable and the different allocations of points to the two sets $S_1$ and $S_2$ is highlighted in the stereogram in Figure \ref{fig4}(c). In addition, the approximate $E$-optimal design for the quadratic polynomial model in three variables on the unit ball derived by \citet{detteg:14} is the subset design $\frac{9}{17} S_0+\frac{3}{17} S_1+ \frac{5}{17}S_3$ and thus a weighted form of the central composite design. The stereogram of the sphere points of the design and their relative weighting is shown in Figure \ref{fig4}(d). Note that  the design itself is rather curious in that more than half of the weight is placed on the centre point.
\bigskip

\noindent{\bf 4.2 Geometric Isomorphism}

\citet{gt:12}, in Example 2 of their paper, considered optimal designs which accommodate inference for a full quadratic polynomial model in three independent variables with support points on the surface and at the centre of a ball.
The authors focussed on nonsingular designs with 18 support points taken, with replacement, from the $27$ individual points belonging the subsets  $S_0, S_1, S_2$ and $S_3$ of the previous subsection. 
More specifically, they considered, {\em inter alia}, D$_S$-optimal designs which minimise the volume of the confidence region for the parameters of the model with the intercept treated as a nuisance parameter and  (DP)$_S$-optimal designs which minimise the D$_S$-criterion multiplied by an appropriate quantile of the $F$-distribution raised to a power equal to the number of parameters of interest, in this case nine. However, over $3 \times 10^{10}$ possible designs must be enumerated in order to identify those designs which are strictly D$_S$- and (DP)$_S$-optimal. As a consequence, \citet{gt:12} used a heuristic, the exchange algorithm, which is documented in, for example,  \citet{adt:07}, to search for such designs.

The question now arises as to how many strictly D$_S$- and (DP)$_S$-optimal designs are generated by the exchange algorithm and, if more than one, how the resultant designs are related. This question in turn impacts on the performance of the algorithm. To this end, the exchange procedure of  \citet{gt:12}  was run five million times in the programming language \citet{gauss:24} and the requisite optimal designs which were distinct identified. Eight distinct D$_S$-optimal designs with two centre points and $1,080$ distinct (DP)$_S$-optimal designs, with $216$ including a single centre point and $864$ two centre points, were so generated. The designs were then examined in the context of isomorphism and, since the factors $0, 1, \frac{1}{\sqrt{2}}$ and $\frac{1}{\sqrt{3}}$ are quantitative, in that of geometric isomorphism \citep{chengy:04}. Specifically, two designs are said to be geometrically isomorphic if the designs can be obtained, one from the other, by rotations and reflections within the design space. In the present example, the isomorphic properties of the designs are determined solely by the spherical points and the actions of rotation and reflection can therefore be mapped directly onto the appropriate stereograms. 

\begin{figure}[htb!]
\vspace{-15.5mm}
\centering

\begin{subfigure}[b]{0.495\textwidth}
\includegraphics[width=\textwidth]{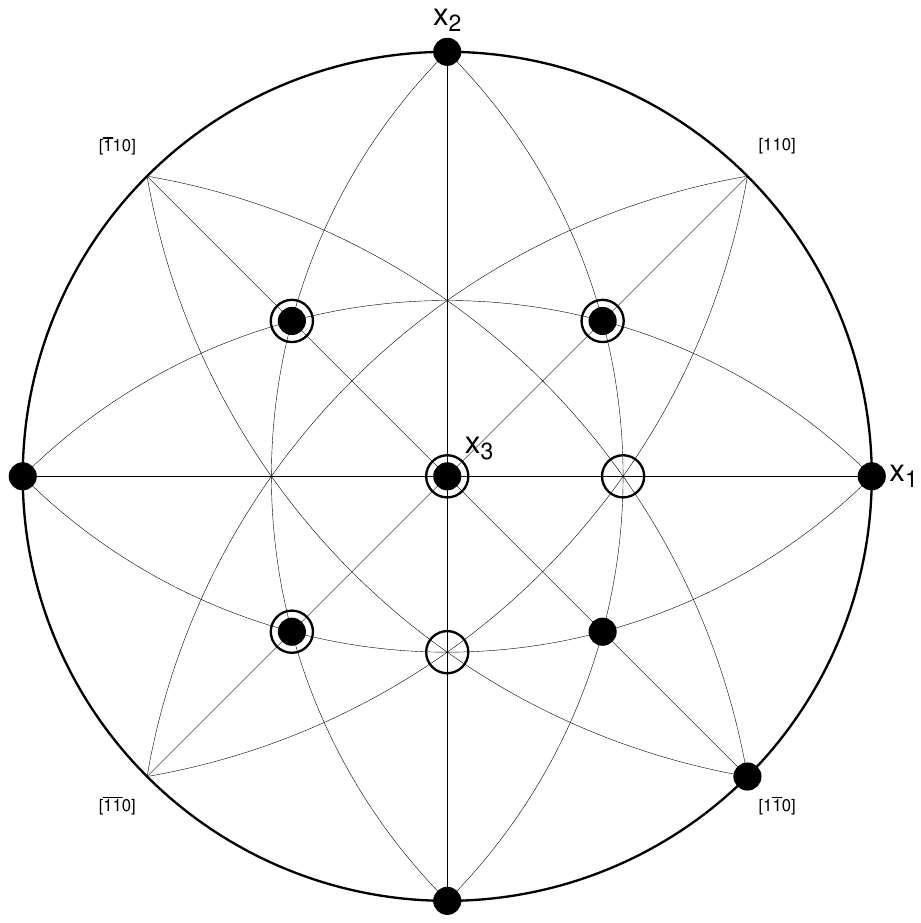}
\vspace{-35mm} \caption{Design of Gilmour and Trinca (2012)}
\end{subfigure}
\hfill
\begin{subfigure}[b]{0.495\textwidth}
\includegraphics[width=\textwidth]{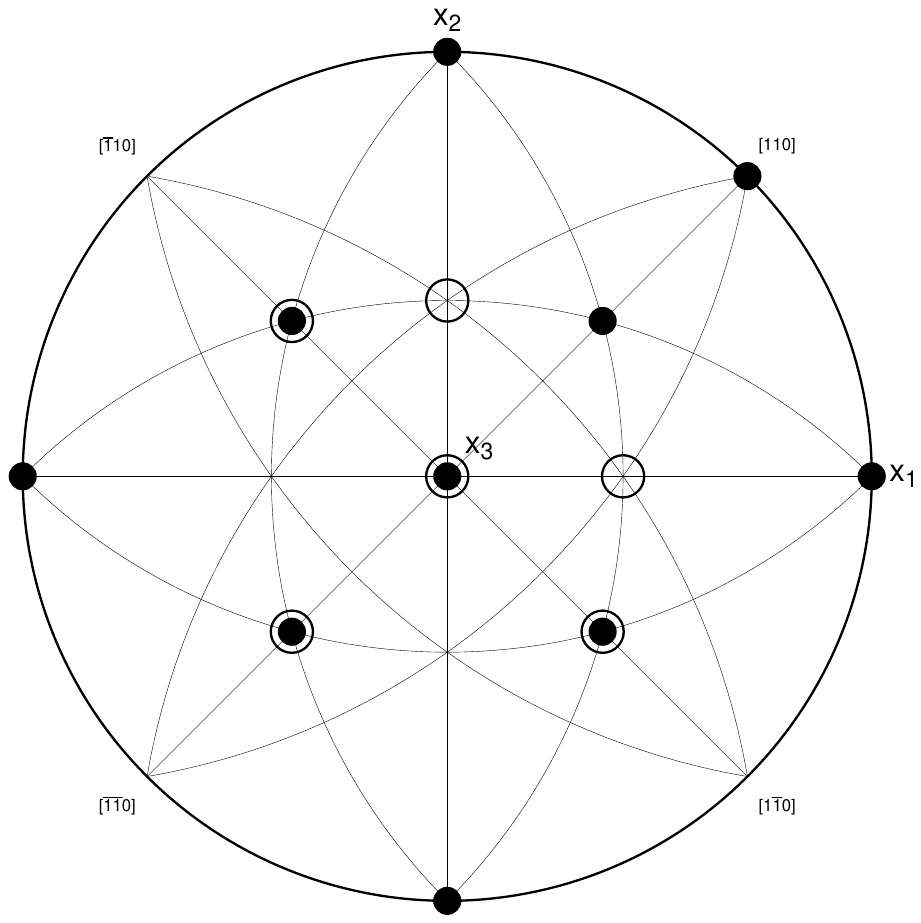} 
\vspace{-35mm} \caption{Rotate (a) through $\displaystyle -\frac{\pi}{2}$}
\end{subfigure}

\vspace{0mm}

\caption{Stereograms of exact isomorphic D$_S$-optimal designs for (a) the design of Gilmour and Trinca (2012) and (b) the design obtained  from (a) by an  anticlockwise $90^{\circ}$ rotation around  the $x_3$ axis. }
\label{fig5}
\end{figure}

The D$_S$-optimal design given in Table 2 of \citet{gt:12} comprises 16 points on the sphere, none of which are replicated, and two centre points and the stereogram of the spherical points of the design is presented in Figure \ref{fig5}(a). It is immediately clear that a further three designs can be generated from the design by repeated anticlockwise rotations through $90^{\circ}$ about the $x_3$ axis and, to illustrate, a stereogram of one of these is shown in Figure \ref{fig5}(b). A further four designs can then be obtained by reflecting these designs  through the plane perpendicular to the $x_3$ axis, that is the $(x_1,x_2)$-plane.  The resultant eight designs so obtained are geometrically isomorphic and therefore D$_S$-optimal and account for the eight distinct D$_S$-optimal designs generated by the exchange algorithm.  

\begin{figure}
\vspace{-25mm}

\centering
\begin{subfigure}[b]{0.475\textwidth}
\includegraphics[width=\textwidth]{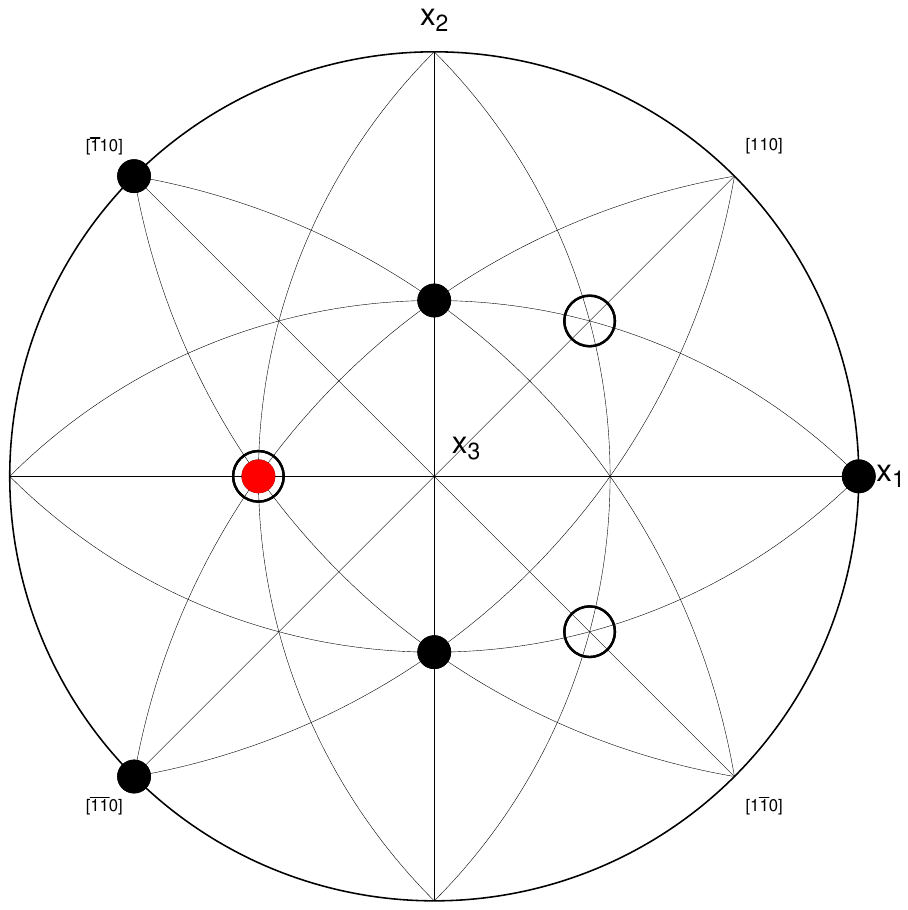} 
\vspace{-35mm} \caption{(DP)$_S$-optimal design(1)}
\end{subfigure}
\hfill
\begin{subfigure}[b]{0.475\textwidth}
 \includegraphics[width=\textwidth]{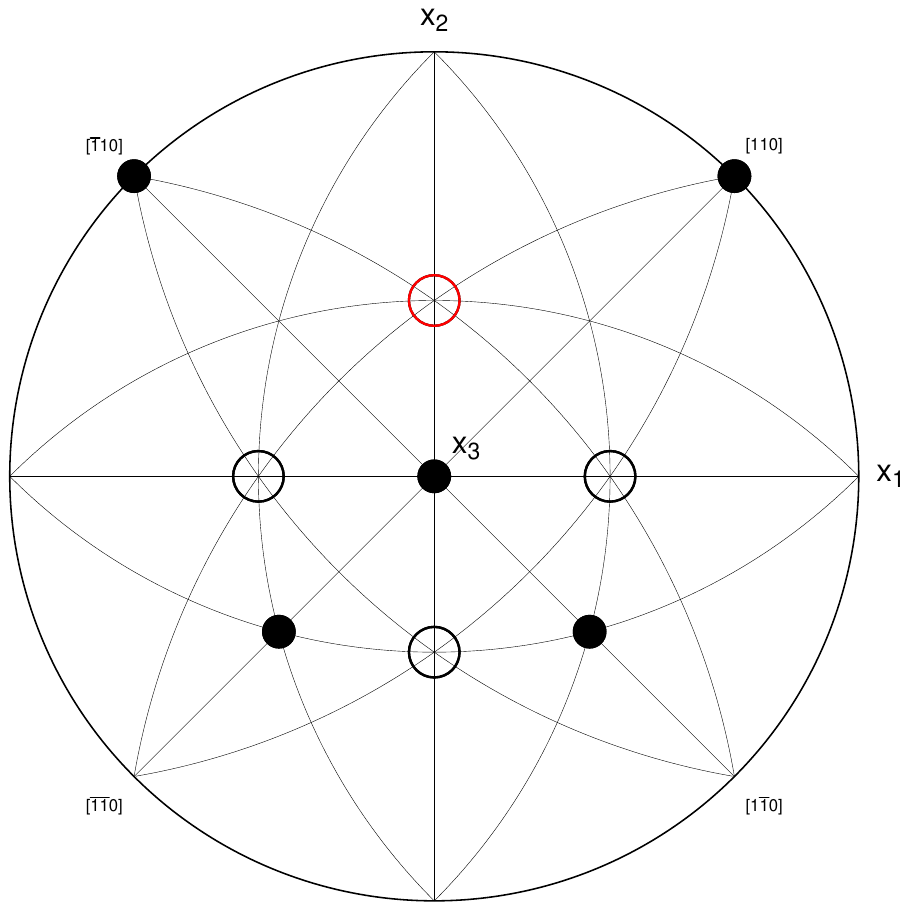} 
\vspace{-35mm} \caption{(DP)$_S$-optimal design(2)}
\end{subfigure}
\hfill
\begin{subfigure}[b]{0.475\textwidth}
 \includegraphics[width=\textwidth]{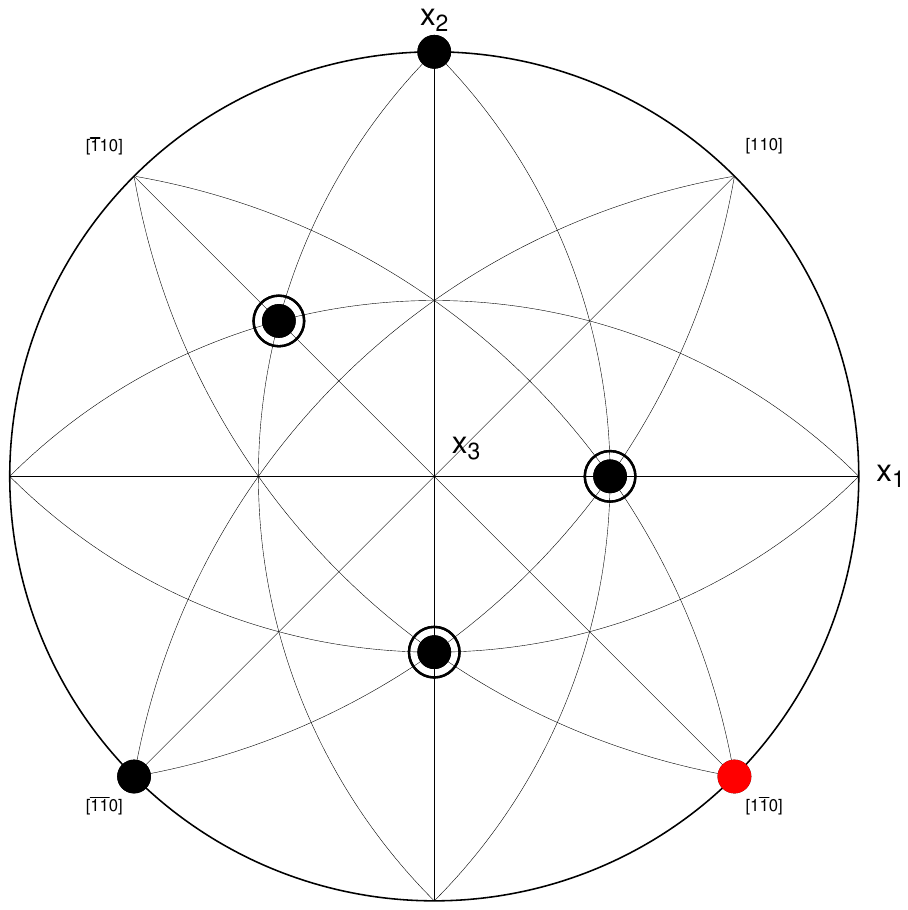} 
\vspace{-35mm} \caption{(DP)$_S$-optimal design(3)}
\end{subfigure}
\hfill
\begin{subfigure}[b]{0.475\textwidth}
\includegraphics[width=\textwidth]{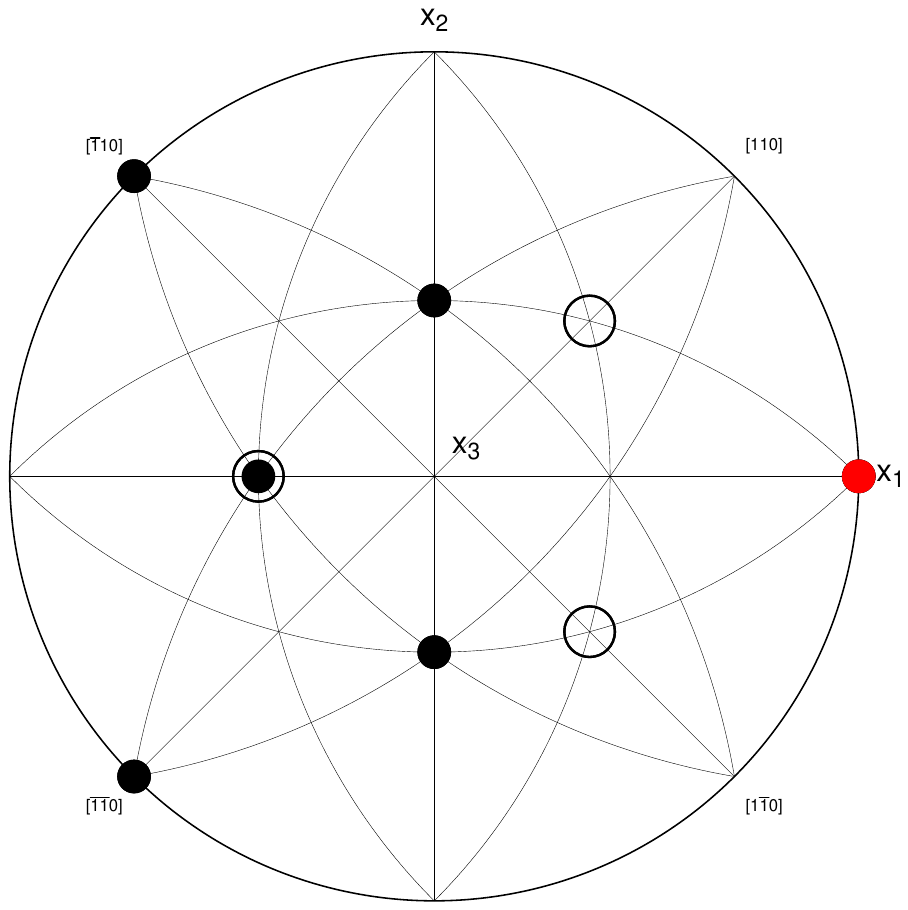} 
\vspace{-35mm} \caption{(DP)$_S$-optimal design(4)}
\end{subfigure}
\vspace{10mm}

\caption{A stereogram of the exact (DP)$_S$-optimal design of Gilmour and Trinca (2012) is shown in (a) and stereograms of two exact (DP)$_S$-optimal designs, which exhibit different patterns of points but which are geometrically isomorphic to (a), are shown in (b) and (c).  The stereogram of a (DP)$_S$-optimal design which is not isomorphic to (a) is shown in (d). Points which occur once are indicated in red and points which occur twice in black.}
\label{fig6}
\end{figure}

The problem of identifying (DP)$_S$-optimal designs which are geometrically isomorphic  from the 1,080 optimal designs generated by the exchange algorithm is a little more nuanced than that relating to the D$_S$-optimal designs. Consider first the design reported in Table 2 of \citet{gt:12} which is one of the 216  (DP)$_S$-optimal designs with a single centre point. 
This $18$-point design is based on the one centre point and on the set of nine points on the sphere
$$
B = \left \{(100), (\frac{1}{\sqrt{3}}, \pm \frac{1}{\sqrt{3}}, -\frac{1}{\sqrt{3}}), 
(0, \pm \frac{1}{\sqrt{2}}, \frac{1}{\sqrt{2}}),
 (- \frac{1}{\sqrt{2}}, 0, \pm \frac{1}{\sqrt{2}}),(-\frac{1}{\sqrt{2}}, \pm \frac{1}{\sqrt{2}}, 0) \right \},
 $$
termed here the basis set, with each point replicated twice except for the point $(-\frac{1}{\sqrt{2}}, 0, \frac{1}{\sqrt{2}})$ which occurs once.
A stereogram of the 17 points on the sphere of the design is presented in Figure \ref{fig6}(a), with the point occurring once indicated in red. 
Two designs can now be obtained by rotation twice through $180^{\circ}$ about the axis defined by the vector $[111]$ and are geometrically isomorphic to the original design of \citet{gt:12}.  To illustrate, the stereograms of the two resultant designs are shown in Figures \ref{fig6}(b) and \ref{fig6}(c).  As an aside,  the designs can also be constructed by cycling the original axes $(x_1, x_2, x_3)$ of \citet{gt:12} twice as $(x_2, x_3, x_1)$ and $(x_3, x_1, x_2)$.
Eight geometrically isomorphic designs can now be generated from each of the three designs displayed in Figures \ref{fig6}(a), (b) and (c) by rotation through $90^{\circ}$  about the $x_3$-axis and reflection about an appropriate plane  to yield a total of $3 \times 8=24$ designs.
The design of \citet{gt:12} is therefore representative of a class of 24 geometrically isomorphic designs.
\bigskip

An examination of the results from the exchange algorithm indicated that designs with one centre point and any one of the nine points in the basis set $B$ occurring once and not twice are (DP)$_S$-optimal. In terms of geometric isomorphism, consider now a design  constructed directly  from  that of \citet{gt:12} but with the the point $(-\frac{1}{\sqrt{2}},0,\frac{1}{\sqrt{2}})$ now replicated twice and the point $(1,0,0)$ occurring once. For clarity, a stereogram of the design is shown in Figure \ref{fig6}(d), with the point occurring once indicated in red. It is clear that the design so constructed is not geometrically isomorphic to the design of \citet{gt:12}. However, by using arguments based on rotations and reflections, it is straightforward to show that the design represents a class of 24 isomorphic (DP)$_S$-optimal designs. Furthermore, the same result holds for all designs constructed with one of the points in the set $B$ occurring once. There are therefore nine classes of designs, each comprising 24 geometrically isomorphic designs, and thus a total of $9 \times 24= 216$ designs.

Consider now the $864$ (DP)$_S$-optimal designs with two centre points generated from multiple runs of the exchange algorithm.  A cursory examination of the results indicated that the points on the sphere of the design comprise two distinct points from the basis set $B$ chosen to occur once and the seven remaining points in $B$ replicated twice. The $\binom{9}{2} = 36$ designs so defined were then constructed independently and shown to be (DP)$_S$-optimal.  
It now follows from arguments similar to those presented for the (DP)$_S$-optimal designs with one centre point that each of the  designs so constructed represents a  class of $24$ geometrically isomorphic designs, thereby yielding $36 \times 24=864$ designs which are (DP)$_S$-optimal and accounting for the designs generated by  the exchange algorithm.

This example serves to illustrate the the importance of examining all optimal designs generated routinely by an exchange algorithm.
\bigskip
\bigskip

\noindent{\bf 5. Discussion}

The focus of the present study is on the use of the stereographic projection within the context of design in order to map points on the three-dimensional unit sphere onto the stereogram in two dimensions. The stereogram gives a clear visualisation of the spatial arrangement of the points on the sphere, as demonstrated here within the context of spherical $t$-designs and regression modelling. 
There are, in fact, many ways of representing an arrangement of points on the three-dimensional sphere in two dimensions, that is on paper, other  than the stereogram. Thus, most simply, a drawing of the sphere with the points added or a sketch of the solid with vertices defined by the points can be used but these approaches are tedious to implement and not helpful in comparative studies. Alternatively, since a set of  points on the sphere defines a convex polytope, the points can be represented by a 3-vertex-connected planar graph, that is a polyhedral graph, but the sense of the spatial arrangement is then lost. There are, in addition,  projections, other than the stereographic projection, such as the orthographic and the gnomonic projections, which can be invoked but these distort the symmetry and, in the latter case, projects the points onto an infinite plane tangential to the sphere. On balance therefore, none of the approaches discussed here offers the ease of use, the sense of the spherical space and the flexibility provided by the stereogram, as is required within the context of design. 

It would be remiss not to mention the use of computer-generated three-dimensional models which represent points and vertices of solids on the sphere and which are available in, for example, the website devoted to spherical $t$-designs of \citet{graefweb} and  the programming language \citet{math:24}. Such models can be rotated in space and thereby examined but it is nevertheless not easy to identify the inbuilt symmetry. The question also arises as to whether points on the four-dimensional sphere can be represented as three- or two-dimensional objects. There are ways of doing this, as  for example with a Schlegel diagram, but these are complicated to interpret and not immediately  useful in design.

The applications presented in this study are of interest in themselves. Spherical $t$-designs are researched extensively from a theoretical perspective in algebraic combinatorics and find application in technologies such as ambisonics. In addition, the designs are often cited, albeit somewhat peripherally, within the context of  quantum theory. Designs for regression models with design space the unit ball are of interest in the area of design of experiments and, more specifically, in optimal design but tend to be represented, at best, as points  on three-dimensional spheres. Stereograms can therefore be used to great advantage in research papers and, more particularly, in text books on experimental design to represent the spherical components of designs on the ball. 

\bigskip
\noindent{\bf Acknowledgements}  The author would like to thank the University of Cape Town and the National Research Foundation (NRF) of South Africa, grant (UID) 119122, for financial support. Any opinion, finding and conclusion or recommendation expressed in this material is that of the author and the NRF does not accept liability in this regard.

\bibliographystyle{chicago}	   
\bibliography{stereograms}
\end{document}